\documentclass[11pt]{article}

\usepackage[preprint]{acl}
\usepackage{pifont}

\usepackage{times}
\usepackage{latexsym}
\usepackage[T1]{fontenc}
\usepackage[utf8]{inputenc}
\usepackage{microtype}
\usepackage{amssymb}
\usepackage{inconsolata}
\usepackage{bbding}
\usepackage{pifont}
\usepackage{wasysym}
\usepackage{amssymb}
\usepackage{graphicx}
\usepackage{booktabs}   
\usepackage{url}
\usepackage{fontawesome5}
\usepackage{xcolor}
\usepackage{tikz}
\newcommand{\yes}{\faCircle}
\newcommand{\partially}{\faAdjust}
\newcommand{\no}{\textcolor{black!35}{\faCircle[regular]}}
\newcommand{\rext}{\faPuzzlePiece}   
\newcommand{\rdev}{\faCode}          
\newcommand{\rnprog}{\faUser}        

\title{LumiXAI: A Modular Full-Stack Framework for Feature Attribution}

\author{
Alfio Ferrara$^{1}$ \and
Lorenzo Gatta$^{2}$ \and
Sergio Picascia$^{2}$ \and
Elisabetta Rocchetti$^{2}$ \\
\\
$^{1}$Department of Literary Studies, Philology and Linguistics, Università degli Studi di Milano, Milan, Italy \\
$^{2}$Department of Computer Science, Università degli Studi di Milano, Milan, Italy \\
\\
\texttt{alfio.ferrara@unimi.it, lorenzo.gatta@studenti.unimi.it}\\
\texttt{sergio.picascia@unimi.it, elisabetta.rocchetti@unimi.it}
}
\begin{document}
\maketitle

\begin{abstract}
Feature attribution is a central tool of model interpretability, yet the software through which it is applied remains fragmented: individual tools specialize along narrow axes, such as a single modality, a code API or a GUI, or a fixed rather than extensible method set, and rarely combine these strengths. Moreover, many explainability tools are designed primarily for domain experts, requiring programming skills or familiarity with attribution methods that can make them difficult for non-expert users to access. In this article, we present \textbf{LumiXAI}, a modular full-stack framework that consolidates attribution analysis into a single system. It couples classification and generative attribution with an interactive GUI supporting bidirectional exploration, a plug-in architecture for registering new models and methods, and three access tiers serving non-programmers, developers, and extenders from one backend. Its contribution is a system that operationalises established attribution methods under one interface, one interaction model, and one persistence layer, with containerised services and persistent results making analyses reproducible across machines.

\end{abstract}

\section{Introduction}
\label{sec:intro}

As machine learning models are increasingly deployed in consequential settings, understanding \emph{why} a model produces a given output has become as important as the output itself. Feature attribution addresses this by assigning importance scores to input elements based on their contribution to a prediction or generation. 
As attribution has grown into a broad methodological family spanning gradient-based, perturbation-based, and attention-based techniques, a practical question arises: how can practitioners navigate, explore, and systematically evaluate these methods?


The tooling through which attribution is applied has developed unevenly across the field: individual tools tend to specialize along a narrow set of axes, such as scope, kind of interaction, or extensible method set, and rarely combine these strengths altogether. A researcher who needs attribution across several model types and interaction styles must therefore assemble and reconcile multiple tools with incompatible interfaces, output formats, and deployment assumptions, an integration burden that impedes precisely the reproducible, exploratory analysis that attribution is meant to support.

\begin{figure}[t]
    \centering
    \includegraphics[width=\linewidth]{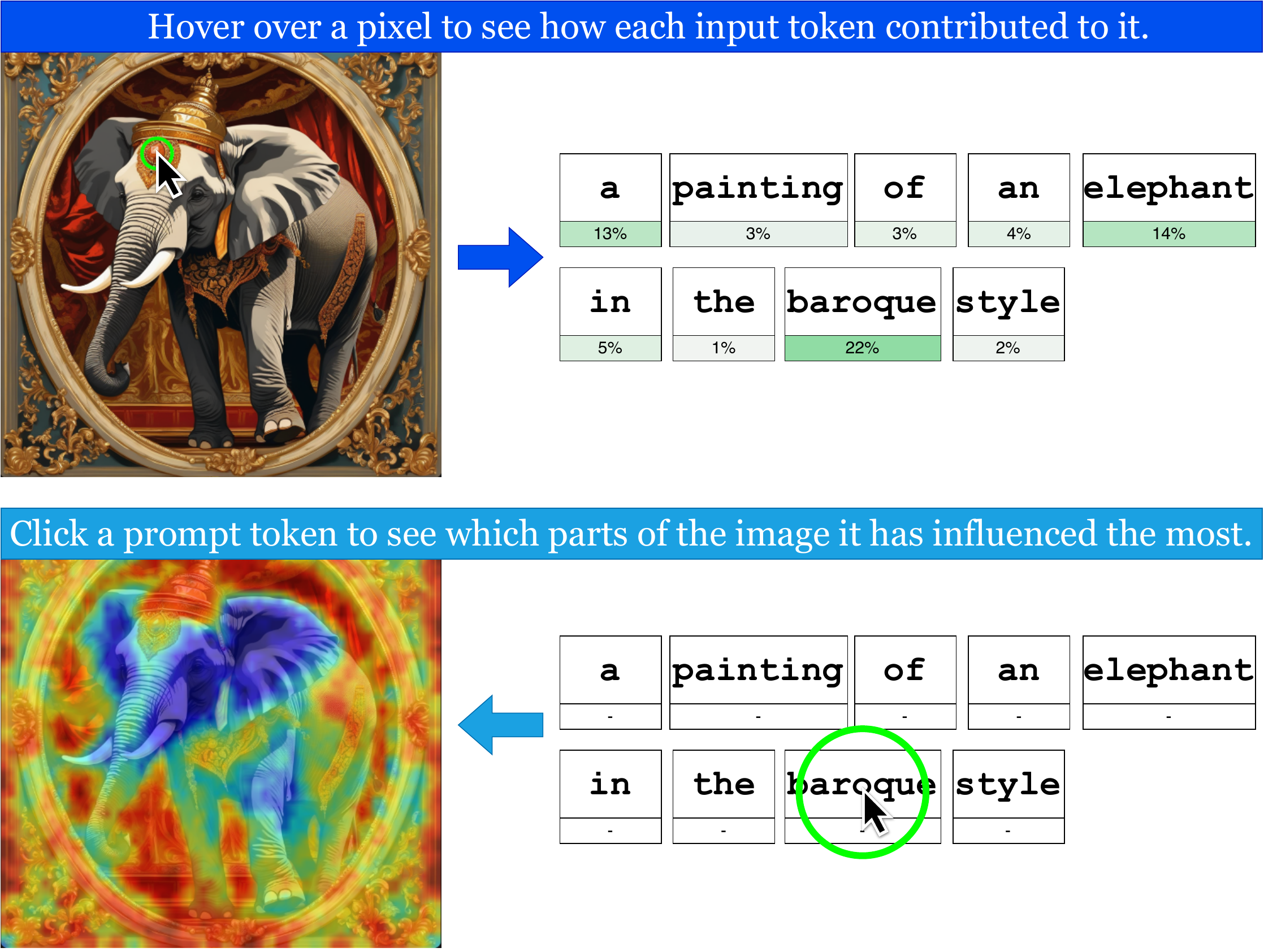}
    \caption{Bidirectional exploration of text-to-image attribution in LumiXAI.
    \emph{Top} (image$\rightarrow$token): hovering a pixel reads that point from every prompt token's attention matrix and recolors the prompt cards.
    \emph{Bottom} (token$\rightarrow$image): clicking a prompt token overlays its attribution heatmap on the generated image.}
    \label{fig:intro}
\end{figure}

This paper presents \textbf{LumiXAI}\footnote{Code and documentation are available at \url{https://github.com/TechLory/LumiXAI}; an online demo is available at \url{http://lumixai.islab.di.unimi.it/}.}, a modular, full-stack framework for feature-attribution analysis across text and image classification, autoregressive text generation, and text-to-image diffusion. LumiXAI is designed to make attribution analyses inspectable, reproducible, and extensible within a single workflow: model families and attribution methods are exposed as registered plug-ins; each layer runs as an independent Docker service; analysis metadata and attribution outputs are persisted for later inspection and sharing; and the same backend supports a web GUI, a Python SDK, and a smart-batch mode for larger experiments. In this way, LumiXAI brings together interaction styles, user groups, and model families that existing tools typically address only in isolation. Figure~\ref{fig:intro} illustrates an example of the interaction with LumiXAI for text-to-image attribution. In addition to the conventional token-to-image view, where a selected prompt token highlights the regions it contributes to, LumiXAI supports the reverse direction: users can hover over an image location and inspect which prompt tokens are most responsible for that point.

The remainder of the paper is organised as follows. Section~\ref{sec:related} situates LumiXAI with respect to existing attribution methods and interpretability toolkits. Section~\ref{sec:system} describes the framework architecture, including its plug-in interface, service-based deployment, storage layer, and user-facing access modes. Section~\ref{sec:cases} presents case studies illustrating LumiXAI across supported model families and interaction scenarios. Section~\ref{sec:evaluation} reports a user evaluation of the system's usability and interpretability workflow. Section~\ref{sec:conclusion} concludes with current limitations and directions for future work.



\section{Related Work}
\label{sec:related}

Attribution tooling has proliferated alongside interpretability research, but existing tools tend to specialise along a small number of axes and rarely combine their strengths. Table~\ref{tab:comparison} places a representative selection along the five dimensions that structure the discussion below: the input modalities with first-class attribution support (text classification, generative text, image classification, and text-to-image generation); whether attribution is defined over the generative decoding process rather than a single classification output; whether the tool ships an interactive GUI; whether its architecture is designed to accommodate new methods or model types; and the audiences it serves. We distinguish three such audiences, since few tools address more than one: \emph{extenders} (\rext), who implement a new attribution method or adapt an unsupported model type and need an interface in which to exercise it; \emph{developers} (\rdev), who apply existing methods through a code API; and \emph{non-programmers} (\rnprog), who work entirely through a GUI.

\begin{table}[t]
\centering
\scriptsize
\setlength{\tabcolsep}{2pt}
\begin{tabular}{l l c c c c}
\toprule
\textbf{Tool} & \textbf{Scope} & \textbf{Gen. Attr.} & \textbf{GUI} & \textbf{Ext.} &  \textbf{Audience} \\
\midrule
Captum             & TC, TG, IC & \yes       & \no  & \yes   & \rext\,\rdev \\
SHAP / LIME        & TC                & \no        & \no  & \no    & \rdev \\
OmniXAI            & TC, IC                & \no        & \yes & \no   & \rdev\,\rnprog \\
AIX360             & TC, IC                & \no        & \no  & \no    & \rdev \\
AllenNLP Interpret & TC, TG    & \partially & \yes & \yes   & \rext\,\rdev \\
LIT                & TC, TG      & \partially & \yes & \yes   & \rext\,\rnprog \\
ecco               & TG           & \yes       & \yes & \no    & \rdev \\
Thermostat         & TC                & \no        & \no  & \no    & \rdev \\
Inseq              & TG           & \yes       & \no  & \yes   & \rext\,\rdev \\
ferret             & TC                & \no        & \no  & \yes   & \rext\,\rdev \\
CafGa              & TG           & \yes       & \yes & \no    & \rnprog \\
ICX360             & TG           & \yes       & \no  & \no   & \rdev \\
\textbf{LumiXAI} & \textbf{TC, TG, IC, T2I} & \yes & \yes  & \yes & \rext\,\rdev\,\rnprog \\
\bottomrule
\end{tabular}
\caption{Comparison of attribution tools. \emph{Scope}: TC~=~text classification, TG~=~text generation, IC~=~image classification, T2I~=~text-to-image generation. \emph{Generative attribution}: \yes~first-class attribution over the generative decoding process (autoregressive or diffusion); \partially~limited to masked-token or single-step prediction, or salience confined to particular seq2seq settings; \no~classification or regression outputs only. \emph{GUI}: whether it ships an interactive visual interface, e.g. standalone application or interactive notebook widget. \emph{Ext.}: if it is designed for adding new attribution methods or model types. \emph{Audience}: \rext~extender (adds methods or models), \rdev~developer (applies methods through a code API), \rnprog~non-programmer (works through the GUI alone). Cells reflect released versions as of July~2026 (Captum~0.9.0, SHAP~0.52.0, Inseq~0.7.1, LIT~1.3.1, CafGa~0.0.6, ICX360~0.1.0).}
\label{tab:comparison}
\end{table}


General-purpose libraries provide broad attribution backends but limited interactive support. LIME~\citep{ribeiroWhyShouldTrust2016c} and SHAP~\citep{lundberg2017unified} offer model-agnostic explanations and plotting utilities, but are mainly used through code and are classification-oriented. Captum~\citep{kokhlikyanCaptumUnifiedGeneric2020} provides an extensible PyTorch library of gradient- and perturbation-based methods, now including support for image masks and language-model attribution~\citep{miglaniUsingCaptumExplain2023}, but its interactive companion, Captum Insights, was retired after v0.8.0. OmniXAI~\citep{yangOmniXAILibraryExplainable2022} and AIX360~\citep{aryaOneExplanationDoes2019} broaden coverage across data types, yet remain centered on classification rather than generative decoding.
NLP-specific tools offer richer model inspection but cover only parts of our target setting. AllenNLP Interpret~\citep{wallaceAllenNLPInterpretFramework2019} exposes saliency and gradient primitives with reusable front-end components, but is tied to the deprecated AllenNLP framework. LIT~\citep{tenneyLanguageInterpretabilityTool2020} provides an extensible browser-based workbench for classification and sequence-to-sequence models, although it is broader than attribution alone. ecco~\citep{alammarEccoOpenSource2021a} visualizes language-model behavior interactively from code, while Thermostat~\citep{feldhusThermostatLargeCollection2021} distributes precomputed classifier explanations. ferret~\citep{attanasioFerretFrameworkBenchmarking2023} complements these systems by benchmarking attribution methods for transformer classifiers.
The closest tools to LumiXAI target generative text. Inseq~\citep{sartiInseqInterpretabilityToolkit2023} supports feature attribution for decoder-only and encoder-decoder generation, but exposes results through a Python API and does not address image models. ICX360~\citep{weiICX360InContextEXplainability2025} attributes LLM generations to their input context, but is likewise text-only and API-driven. CafGa~\citep{boyleCafGaCustomizingFeature2025} adds interactive controls for attribution granularity, including Jupyter widgets, but remains confined to text.


LumiXAI differs in providing the widest scope support, combining generative attribution across text and text-to-image diffusion with an interactive GUI, documented extension interfaces, and multiple access tiers backed by the same services. This design lets non-programmers inspect analyses visually, developers access them programmatically, and extenders add new model families or attribution methods through the same registry.

\section{The LumiXAI Framework}
\label{sec:system}


LumiXAI is a client--server system that isolates model-specific and compute-intensive functionality behind a REST API. Clients interact with this boundary rather than with model classes directly, allowing the same backend to support the web interface, Python SDK, and batch workflows. Internally, the attribution engine is organised around two abstract plug-in interfaces, wrappers and attributors, which decouple model access from explanation computation and make it possible to add new tasks, models, or attribution methods without changing the core system.

\subsection{Architecture Overview}
\label{subsec:arch}



\begin{figure}[t]
  \centering
  \includegraphics[width=\linewidth]{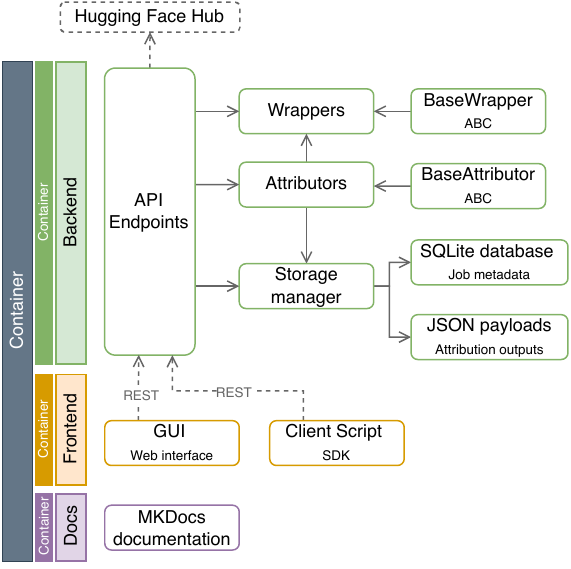}
  \caption{LumiXAI architecture. The backend exposes a REST API, orchestrates wrapper and attributor plug-ins, retrieves models from the Hugging Face Hub, and stores metadata in SQLite with attribution payloads as JSON files. Backend, frontend, and documentation are deployed as independent containers, and both the web GUI and client scripts consume the same API.}
  \label{fig:architecture}
\end{figure}

Figure~\ref{fig:architecture} summarizes the system architecture. LumiXAI is deployed as three independent services: a FastAPI backend, a Next.js frontend, and MkDocs documentation. The backend exposes the REST API, hosts the attribution engine, and is the only component that loads model weights. Wrappers retrieve and adapt models, currently from the Hugging Face Hub, while attributors compute explanations through a shared interface. 

The frontend and SDK are both ordinary API clients and contain no attribution logic. Attribution requests may be long-running, so the backend handles them as jobs: each request receives an identifier, while computation proceeds asynchronously. Access to the accelerator is serialised to avoid memory conflicts, and device placement is resolved at load time across CUDA, Apple MPS, or CPU. 

Completed analyses are persisted through a hybrid storage layer. Structured job metadata, such as model, method, prompt, status, and elapsed time, are stored in SQLite. Larger attribution payloads, including token scores, per-step generation traces, attention matrices, and generated images, are serialised as JSON files referenced from the corresponding database row. This design keeps analyses available for later inspection, export, or transfer.

\subsection{Extensibility: Wrappers and Attributors}
\label{subsec:ext}


Extensibility is enforced by keeping models and attribution algorithms behind abstract interfaces. A \texttt{BaseWrapper} provides a uniform view of a model family: it loads weights, places the model on the selected device, and exposes the forward operation. When available, it may also expose embeddings for gradient-based methods, but this is not required for attribution methods that rely on other signals, such as attention.


A \texttt{BaseAttributor} defines an attribution method as a mapping from a wrapped model and an input to a standardised output object. The wrapper is injected into the attributor at construction time, and each attributor declares the wrapper families it supports. The backend uses this declaration to reject incompatible model--method pairings before loading weights, while the frontend uses the same information to disable invalid choices.


Wrappers and attributors are registered as plug-ins. Adding a new model family or attribution method therefore requires one subclass, an applicability declaration, and one registry entry. The client remains independent of these details: it populates its selectors from a backend manifest, and the backend resolves the appropriate wrapper from the task metadata declared by the selected model.


The data contracts crossing the plug-in boundary are modality-agnostic. The atomic unit of input is an \emph{input feature}, i.e. a token with its position for text, a patch or a superpixel for an image, and every attributor returns the same output object irrespective of what was explained. The coverage reported in Section~\ref{subsec:models} is thus what is currently implemented.

\subsection{Supported Models and Methods}
\label{subsec:models}

The current release ships four wrappers. \texttt{HFTextClassificationWrapper} adapts encoder classifiers and returns class logits with token-level scores. \texttt{HFTextGenerationWrapper} adapts autoregressive models through a step-wise decoding loop, producing an \emph{attribution trace} that links each generated token and probability to attribution scores over the context available at that step. \texttt{HFImageClassificationWrapper} supports vision transformers and CNN classifiers, rendering explanations over the model's preprocessed, de-normalised input. \texttt{HFImageWrapper} adapts diffusion-based text-to-image models and exposes the components needed for cross-attention attribution.


LumiXAI provides ten established attribution methods. Seven are available through Captum for text and image classification: Integrated Gradients~\citep{sundararajan2017axiomatic}, DeepLift~\citep{shrikumar2017learning}, Saliency~\citep{simonyan2014deep}, Input$\times$Gradient, GradientSHAP~\citep{lundberg2017unified}, Occlusion~\citep{zeiler2014visualizing}, and LIME~\citep{ribeiro2016why}. For perturbation methods, the perturbed unit is the token for text, a regular patch grid for image Occlusion, and SLIC superpixels~\citep{achanta2012slic} for image LIME. Two further methods, SmoothGrad~\citep{smilkov2017smoothgrad} and Grad-CAM~\citep{selvaraju2017gradcam}, target vision models. Finally, DAAM~\citep{tang2023daam} attributes text-to-image outputs by recovering per-token spatial maps from diffusion cross-attention.


\subsection{Multi-Tier Access}
\label{subsec:tiers}

A single backend is exposed through three access modes, corresponding to the ``Audience'' of Table~\ref{tab:comparison}.
 
\paragraph{Web GUI (\rnprog).}
The Next.js interface lets users configure, launch, and inspect attribution jobs without writing code. Classification views display token or image heatmaps alongside model predictions. Generative views expose the attribution trace bidirectionally: in text generation, users can move from input tokens to generated-token influence or from a generated token back to its supporting context; in text-to-image generation, they can inspect token heatmaps on the image or query a pixel to see which prompt tokens contributed most. The GUI also includes tutorials and a history view for revisiting completed jobs.
 
\paragraph{Python SDK (\rdev).}
A single self-contained module drives the REST API programmatically, for headless use in any environment able to reach the backend. Documented notebooks accompany it and reproduce the case studies of Section~\ref{sec:cases}. For larger sweeps, \texttt{run\_smart\_batch()} reduces repeated model weight loading by grouping jobs by model and method, so that each set of weights is loaded once per group, and the groups are then \emph{ordered} by a configurable strategy (\texttt{fastest\_first}, \texttt{slowest\_first}, or \texttt{none}) suited to shared and dedicated accelerators respectively. Results are collected by polling and returned in the caller's original job order irrespective of completion order, so that the optimization remains transparent to the calling code.

\paragraph{Framework extension (\rext).}
The third mode addresses researchers whose model or attribution method the framework does not yet support, and who therefore act on the codebase rather than through it. Such a user subclasses \texttt{BaseWrapper} or \texttt{BaseAttributor}, declares the new component in the registry, and obtains a plug-in that the backend orchestrates on equal terms with those shipped by default. Because the manifest is derived from the registry, the new method appears in the GUI selectors and is reachable through the REST API and the SDK immediately.

\subsection{Deployment}
\label{subsec:deploy}

LumiXAI is deployed with Docker Compose. A base \texttt{docker-compose.yml} defines the backend, frontend, and documentation services, while an optional \texttt{docker-compose.gpu.yml} enables GPU execution through the NVIDIA Container Toolkit. Services can be run together or independently: the backend is sufficient for SDK use, the frontend enables interactive analysis, and the documentation can be served separately.

The backend image is built with Poetry from a lockfile, and configuration is parameterised through environment variables with in-file defaults. The Hugging Face cache is mounted as a volume so that model weights persist across restarts. Together with the SQLite metadata store and JSON attribution payloads described above, this makes both the execution environment and completed analyses portable across machines.

Runtime is mainly determined by the selected model and attribution method, including model inference, repeated forward or backward passes, and, where applicable, diffusion or generation steps. LumiXAI adds scheduling, serialization, and retrieval around these computations, but these components are not expected to be the main source of latency; we leave a dedicated measurement of framework overhead to future work.

\section{Case Studies}
\label{sec:cases}

We illustrate LumiXAI usage through two examples covering text classification and text-to-image diffusion. Additional examples on image classification and open-ended text generation are provided in the Appendix~\ref{sec:add-case}.



\paragraph{Text Classification.}
\label{subsec:case-classification}



\begin{figure}[ht!]
    \centering
    \includegraphics[width=0.9\linewidth]{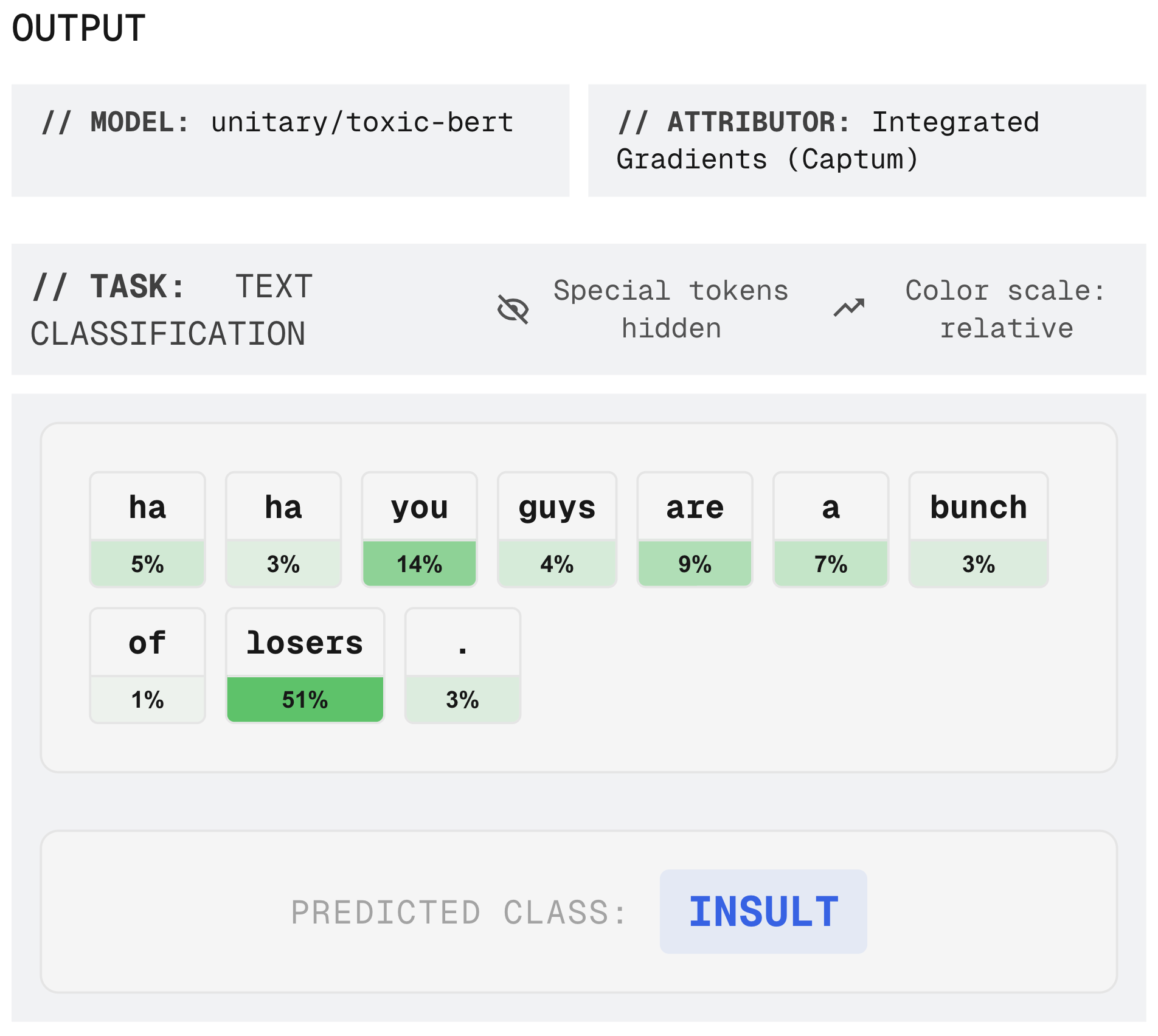}
    \caption{Text classification in LumiXAI. For the comment
\emph{``ha ha you guys are a bunch of losers.''}, classified as
\textsc{insult} by \texttt{unitary/toxic-bert}, tokens are colored by
Integrated-Gradients attribution toward the predicted class; \emph{losers}
dominates the attribution (51\%).}
    \label{fig:case-classification}
\end{figure}

We first examine which input features drive a toxicity classifier. A curated subset of Civil Comments~\citep{borkan2019nuanced} is passed to \texttt{unitary/toxic-bert}, and each prediction is attributed toward the most probable class using Integrated Gradients. In the frontend, tokens are colored by signed attribution score, making it possible to inspect whether the model relies on identity terms, offensive language, or quoted stereotypes when assigning toxicity labels (Figure~\ref{fig:case-classification}).

\paragraph{Text-to-Image Diffusion.}
\label{subsec:case-diffusion}



\begin{figure}[ht!]
\centering
\includegraphics[width=0.9\linewidth]{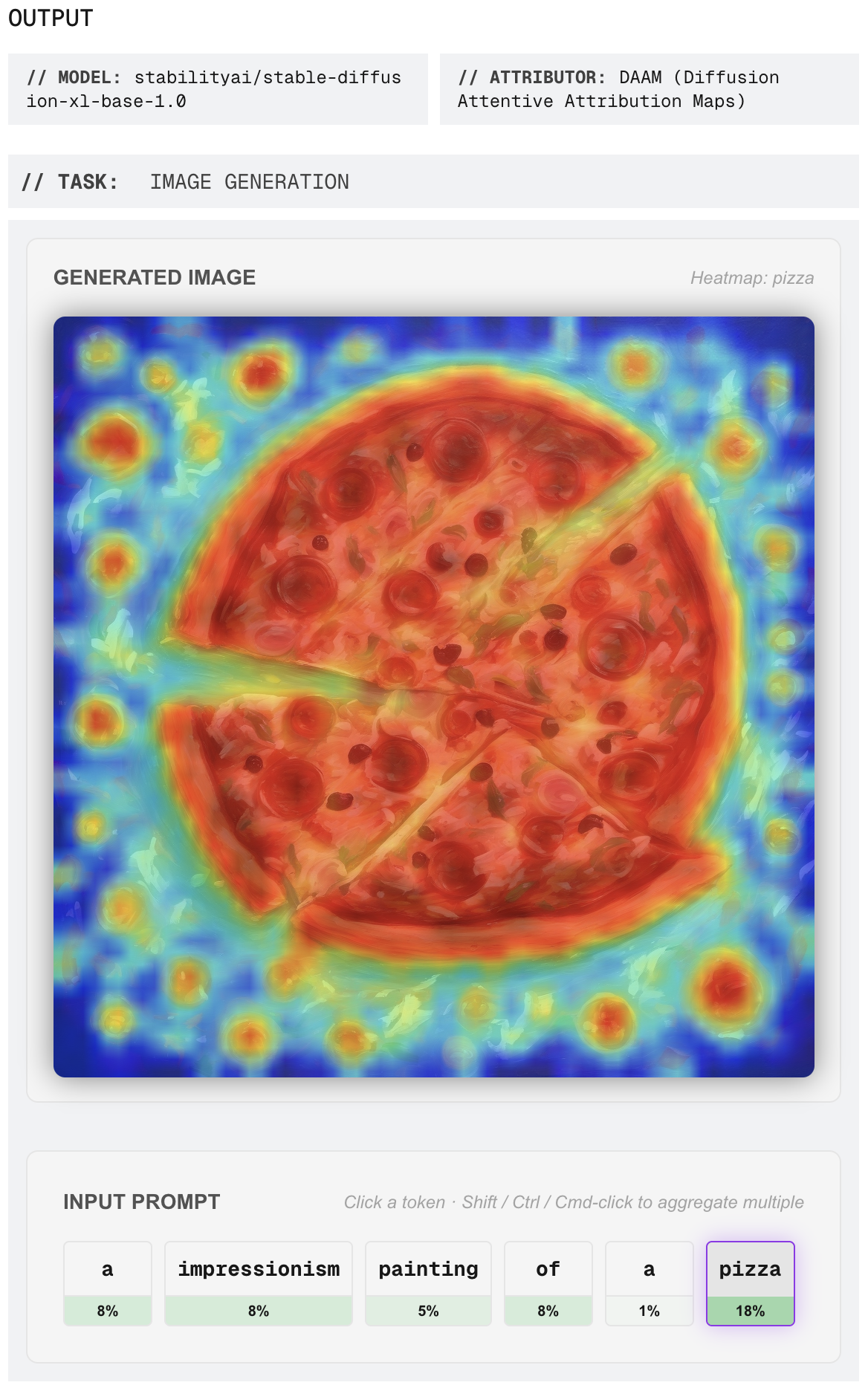}
\caption{Text-to-image attribution in LumiXAI. For the prompt
\emph{``a impressionism painting of a pizza''}, generated by \texttt{Stable
Diffusion XL}, selecting \emph{pizza} overlays its DAAM heatmap
(token$\rightarrow$image), while token percentages report attribution at the
selected pixel (image$\rightarrow$token).}
\label{fig:case-diffusion}
\end{figure}

The second case study explores whether text-to-image diffusion models spatially separate artistic \emph{content} and \emph{style}. Following~\citet{ferrara2025}, we generate images from prompts pairing a subject with an artistic style and attribute pixels to prompt tokens using DAAM. Since diffusion cross-attention links prompt tokens to image regions during denoising, LumiXAI makes it possible to compare the spatial influence of content and style terms directly. Selecting a prompt token overlays its attribution heatmap on the image, while selecting an image region reveals the responsible prompt tokens. This bidirectional view supports inspection of whether content tokens localize to depicted objects, whether style tokens spread across texture and background, and where the model entangles the two (Figure~\ref{fig:case-diffusion}).

\section{User Evaluation}
\label{sec:evaluation}

We conducted a user evaluation to assess whether LumiXAI is usable by different target audiences and whether its interface supports the inspection of attribution results across tasks. The study focuses on usability and workflow effectiveness rather than on the intrinsic quality of the explanations, which depends on the external attribution libraries we employ.

\paragraph{Setup.}
We recruited 12 participants, including 3 participants with experience in machine learning or NLP and 9 non-expert users. Expertise was determined from two background questions on prior experience training or evaluating ML models and using feature-attribution or explainability methods.

Each participant followed the same protocol. After giving consent and answering the background questions, participants completed the LumiXAI tutorials, introducing the supported tasks, attribution views, and interaction patterns, followed by a short comprehension check on the meaning and sign of attribution scores.
Participants then completed 8 tasks across four task types: text classification, image classification, autoregressive text generation, and text-to-image diffusion. For each task type we fixed the model and attribution method and provided the required input (e.g., a text prompt or image); participants launched the attribution jobs and inspected the results, exploring token-level attributions in both directions for text tasks and attribution regions for image tasks. This let us evaluate interface usability while keeping model and method choices controlled across users.
Finally, participants rated usability statements about LumiXAI on a 7-point Likert scale and gave free-text feedback on the most useful feature, on confusing or frustrating aspects, and on anything they wanted to inspect that the system did not allow.

\paragraph{Results.}
Table~\ref{tab:eval} summarizes the questionnaire results. Overall, participants judged the system positively, appreciating in particular the ease of switching between models and attribution methods and its helpfulness in understanding model behaviour, although ratings were consistently higher among expert users. The qualitative feedback confirms this picture, indicating that clearer onboarding, more transparent heatmap colour scaling, and explicit indication of method–model compatibility would be the main avenues for improving accessibility to non-experts.

\begin{table}[ht!]
    \centering
    \small
    \setlength{\tabcolsep}{4pt}
    \begin{tabular}{lccc}
    \toprule
    \textbf{Question} & \textbf{Overall} & \textbf{Experts} & \textbf{Non-experts} \\
    \midrule
    Ease of use        & 4.92 & 5.67 & 4.67 \\
    Learnability       & 5.17 & 6.33 & 4.78 \\
    Enjoyability       & 4.75 & 5.67 & 4.44 \\
    Integration        & 5.33 & 5.67 & 5.22 \\
    Helpfulness        & 5.25 & 6.00 & 5.00 \\
    Reuse intention    & 4.25 & 4.33 & 4.22 \\
    Bidirectionality   & 4.67 & 5.00 & 4.56 \\
    Switching          & 5.58 & 5.67 & 5.56 \\
    Compatibility      & 4.58 & 5.33 & 4.33 \\
    Consistency        & 5.00 & 5.67 & 4.78 \\
    \midrule
    Mean               & 4.95 & 5.53 & 4.76 \\
    \bottomrule
    \end{tabular}
    \caption{Mean user-study ratings (7-point Likert scale) per questionnaire item, overall and by expertise ($n=12$; 3 experts, 9 non-experts).}
    \label{tab:eval}
\end{table}


\section{Conclusion}
\label{sec:conclusion}

We presented \textbf{LumiXAI}, a modular full-stack framework for reproducible and interactive feature-attribution analysis. LumiXAI operationalizes established attribution techniques within a unified system that spans multiple model families, access modes, and user profiles. Through its plug-in architecture, service-based deployment, persistent storage layer, and complementary interfaces, the framework turns attribution workflows that often require ad hoc scripting into analyses that can be inspected, repeated, and extended. The case studies illustrate how LumiXAI supports both targeted debugging and exploratory interpretation across classification, generation, and text-to-image diffusion scenarios. The framework current coverage is necessarily limited to the model families and attribution methods implemented so far, and future work will extend the plug-in library with additional techniques and architectures. In particular, emerging multimodal models, such as image-text-to-text systems, raise new attribution questions that fit naturally within LumiXAI's modality-agnostic design.

\clearpage

\section*{Limitations}
LumiXAI is extensible, but its current implementation covers only a subset of possible tasks and models. Other architectures, especially image-text-to-text vision-language models, require additional wrappers and visualization components. Likewise, the supported attribution methods are representative rather than exhaustive.

The tool also inherits the limitations of attribution methods themselves. Scores can vary with baselines, tokenization, perturbation units, aggregation choices, and preprocessing, and should therefore be treated as diagnostic signals rather than causal explanations. This is especially relevant for generative models, where decoding traces and diffusion attention maps expose useful patterns without fully explaining the underlying computation.

The system was primarily developed and tested under single-user conditions, with concurrency handled only across multiple processes issued by that user. During the user evaluation, some participants encountered bugs caused by multiple users concurrently loading and unloading different configurations on the same backend, a scenario we had not accounted for; robustness under concurrent multi-user load remains future work.

Finally, the user evaluation is limited in scale and scope. It tests whether participants can complete guided attribution tasks and interpret the visualizations, but does not measure long-term use in open-ended research workflows. Moreover, we do not isolate framework overhead from model inference and attribution time, so a dedicated systems benchmark is left to future work.

\section*{Ethics Statement}

Participation in the user evaluation was voluntary, and participants were informed of the study's purpose, procedure, and expected duration before starting. Written informed consent was obtained from all participants prior to the tutorial phase. No personally identifiable information was collected: questionnaire responses and free-text feedback were anonymised at collection time and are reported only in aggregate. Participants could withdraw from the study at any point without providing a reason. The study involved no more than minimal risk, as it consisted of guided interaction with a software interface and did not involve sensitive personal data, deception, or vulnerable populations.


\bibliography{custom}

@inproceedings{alammarEccoOpenSource2021a,
  title = {Ecco: {{An Open Source Library}} for the {{Explainability}} of {{Transformer Language Models}}},
  shorttitle = {Ecco},
  booktitle = {Proceedings of the 59th {{Annual Meeting}} of the {{Association}} for {{Computational Linguistics}} and the 11th {{International Joint Conference}} on {{Natural Language Processing}}: {{System Demonstrations}}},
  author = {Alammar, J},
  editor = {Ji, Heng and Park, Jong C. and Xia, Rui},
  year = 2021,
  month = aug,
  pages = {249--257},
  publisher = {Association for Computational Linguistics},
  address = {Online},
  doi = {10.18653/v1/2021.acl-demo.30},
  urldate = {2026-07-01}
}

@misc{aryaOneExplanationDoes2019,
  title = {One {{Explanation Does Not Fit All}}: {{A Toolkit}} and {{Taxonomy}} of {{AI Explainability Techniques}}},
  shorttitle = {One {{Explanation Does Not Fit All}}},
  author = {Arya, Vijay and Bellamy, Rachel K. E. and Chen, Pin-Yu and Dhurandhar, Amit and Hind, Michael and Hoffman, Samuel C. and Houde, Stephanie and Liao, Q. Vera and Luss, Ronny and Mojsilovi{\'c}, Aleksandra and Mourad, Sami and Pedemonte, Pablo and Raghavendra, Ramya and Richards, John and Sattigeri, Prasanna and Shanmugam, Karthikeyan and Singh, Moninder and Varshney, Kush R. and Wei, Dennis and Zhang, Yunfeng},
  year = 2019,
  month = sep,
  number = {arXiv:1909.03012},
  eprint = {1909.03012},
  primaryclass = {cs.AI},
  publisher = {arXiv},
  doi = {10.48550/arXiv.1909.03012},
  urldate = {2026-07-01},
  archiveprefix = {arXiv}
}

@inproceedings{attanasioFerretFrameworkBenchmarking2023,
  title = {Ferret: A {{Framework}} for {{Benchmarking Explainers}} on {{Transformers}}},
  shorttitle = {Ferret},
  booktitle = {Proceedings of the 17th {{Conference}} of the {{European Chapter}} of the {{Association}} for {{Computational Linguistics}}: {{System Demonstrations}}},
  author = {Attanasio, Giuseppe and Pastor, Eliana and Di Bonaventura, Chiara and Nozza, Debora},
  editor = {Croce, Danilo and Soldaini, Luca},
  year = 2023,
  month = may,
  pages = {256--266},
  publisher = {Association for Computational Linguistics},
  address = {Dubrovnik, Croatia},
  doi = {10.18653/v1/2023.eacl-demo.29},
  urldate = {2026-07-01}
}

@inproceedings{boyleCafGaCustomizingFeature2025,
  title = {{{CafGa}}: {{Customizing Feature Attributions}} to {{Explain Language Models}}},
  shorttitle = {{{CafGa}}},
  booktitle = {Proceedings of the 2025 {{Conference}} on {{Empirical Methods}} in {{Natural Language Processing}}: {{System Demonstrations}}},
  author = {Boyle, Alan David and Cheng, Furui and Zouhar, Vil{\'e}m and {El-Assady}, Mennatallah},
  editor = {Habernal, Ivan and Schulam, Peter and Tiedemann, J{\"o}rg},
  year = 2025,
  month = nov,
  pages = {461--470},
  publisher = {Association for Computational Linguistics},
  address = {Suzhou, China},
  doi = {10.18653/v1/2025.emnlp-demos.32},
  urldate = {2026-07-01},
  isbn = {979-8-89176-334-0}
}

@inproceedings{feldhusThermostatLargeCollection2021,
  title = {Thermostat: {{A Large Collection}} of {{NLP Model Explanations}} and {{Analysis Tools}}},
  shorttitle = {Thermostat},
  booktitle = {Proceedings of the 2021 {{Conference}} on {{Empirical Methods}} in {{Natural Language Processing}}: {{System Demonstrations}}},
  author = {Feldhus, Nils and Schwarzenberg, Robert and M{\"o}ller, Sebastian},
  editor = {Adel, Heike and Shi, Shuming},
  year = 2021,
  month = nov,
  pages = {87--95},
  publisher = {Association for Computational Linguistics},
  address = {Online and Punta Cana, Dominican Republic},
  doi = {10.18653/v1/2021.emnlp-demo.11},
  urldate = {2026-07-01}
}

@misc{kokhlikyanCaptumUnifiedGeneric2020,
  title = {Captum: {{A}} Unified and Generic Model Interpretability Library for {{PyTorch}}},
  shorttitle = {Captum},
  author = {Kokhlikyan, Narine and Miglani, Vivek and Martin, Miguel and Wang, Edward and Alsallakh, Bilal and Reynolds, Jonathan and Melnikov, Alexander and Kliushkina, Natalia and Araya, Carlos and Yan, Siqi and {Reblitz-Richardson}, Orion},
  year = 2020,
  month = sep,
  number = {arXiv:2009.07896},
  eprint = {2009.07896},
  primaryclass = {cs.LG},
  publisher = {arXiv},
  doi = {10.48550/arXiv.2009.07896},
  urldate = {2026-07-01},
  archiveprefix = {arXiv}
}

@inproceedings{miglaniUsingCaptumExplain2023,
  title = {Using {{Captum}} to {{Explain Generative Language Models}}},
  booktitle = {Proceedings of the 3rd {{Workshop}} for {{Natural Language Processing Open Source Software}} ({{NLP-OSS}} 2023)},
  author = {Miglani, Vivek and Yang, Aobo and Markosyan, Aram and {Garcia-Olano}, Diego and Kokhlikyan, Narine},
  editor = {Tan, Liling and Milajevs, Dmitrijs and Chauhan, Geeticka and Gwinnup, Jeremy and Rippeth, Elijah},
  year = 2023,
  month = dec,
  pages = {165--173},
  publisher = {Association for Computational Linguistics},
  address = {Singapore},
  doi = {10.18653/v1/2023.nlposs-1.19},
  urldate = {2026-07-01}
}

@inproceedings{lundberg2017unified,
  title = {A {{Unified Approach}} to {{Interpreting Model Predictions}}},
  booktitle = {Advances in {{Neural Information Processing Systems}}},
  author = {Lundberg, Scott M and Lee, Su-In},
  year = 2017,
  volume = {30},
  publisher = {Curran Associates, Inc.},
  urldate = {2026-07-01}
}

@inproceedings{ribeiroWhyShouldTrust2016c,
  title = {"{{Why Should I Trust You}}?"},
  shorttitle = {"{{Why Should I Trust You}}?},
  booktitle = {Proceedings of the 22nd {{ACM SIGKDD International Conference}} on {{Knowledge Discovery}} and {{Data Mining}}},
  author = {Ribeiro, Marco Tulio and {View Profile} and Singh, Sameer and {View Profile} and Guestrin, Carlos and {View Profile}},
  year = 2016,
  month = aug,
  series = {{{ACM Conferences}}},
  pages = {1135--1144},
  doi = {10.1145/2939672.2939778},
  urldate = {2026-07-01},
  isbn = {978-1-4503-4232-2}
}

@inproceedings{sartiInseqInterpretabilityToolkit2023,
  title = {Inseq: {{An Interpretability Toolkit}} for {{Sequence Generation Models}}},
  shorttitle = {Inseq},
  booktitle = {Proceedings of the 61st {{Annual Meeting}} of the {{Association}} for {{Computational Linguistics}} ({{Volume}} 3: {{System Demonstrations}})},
  author = {Sarti, Gabriele and Feldhus, Nils and Sickert, Ludwig and {van der Wal}, Oskar},
  editor = {Bollegala, Danushka and Huang, Ruihong and Ritter, Alan},
  year = 2023,
  month = jul,
  pages = {421--435},
  publisher = {Association for Computational Linguistics},
  address = {Toronto, Canada},
  doi = {10.18653/v1/2023.acl-demo.40},
  urldate = {2026-07-01}
}

@inproceedings{tenneyLanguageInterpretabilityTool2020,
  title = {The {{Language Interpretability Tool}}: {{Extensible}}, {{Interactive Visualizations}} and {{Analysis}} for {{NLP Models}}},
  shorttitle = {The {{Language Interpretability Tool}}},
  booktitle = {Proceedings of the 2020 {{Conference}} on {{Empirical Methods}} in {{Natural Language Processing}}: {{System Demonstrations}}},
  author = {Tenney, Ian and Wexler, James and Bastings, Jasmijn and Bolukbasi, Tolga and Coenen, Andy and Gehrmann, Sebastian and Jiang, Ellen and Pushkarna, Mahima and Radebaugh, Carey and Reif, Emily and Yuan, Ann},
  editor = {Liu, Qun and Schlangen, David},
  year = 2020,
  month = oct,
  pages = {107--118},
  publisher = {Association for Computational Linguistics},
  address = {Online},
  doi = {10.18653/v1/2020.emnlp-demos.15},
  urldate = {2026-07-01}
}

@inproceedings{wallaceAllenNLPInterpretFramework2019,
  title = {{{AllenNLP Interpret}}: {{A Framework}} for {{Explaining Predictions}} of {{NLP Models}}},
  shorttitle = {{{AllenNLP Interpret}}},
  booktitle = {Proceedings of the 2019 {{Conference}} on {{Empirical Methods}} in {{Natural Language Processing}} and the 9th {{International Joint Conference}} on {{Natural Language Processing}} ({{EMNLP-IJCNLP}}): {{System Demonstrations}}},
  author = {Wallace, Eric and Tuyls, Jens and Wang, Junlin and Subramanian, Sanjay and Gardner, Matt and Singh, Sameer},
  editor = {Pad{\'o}, Sebastian and Huang, Ruihong},
  year = 2019,
  month = nov,
  pages = {7--12},
  publisher = {Association for Computational Linguistics},
  address = {Hong Kong, China},
  doi = {10.18653/v1/D19-3002},
  urldate = {2026-07-01}
}

@misc{yangOmniXAILibraryExplainable2022,
  title = {{{OmniXAI}}: {{A Library}} for {{Explainable AI}}},
  shorttitle = {{{OmniXAI}}},
  author = {Yang, Wenzhuo and Le, Hung and Laud, Tanmay and Savarese, Silvio and Hoi, Steven C. H.},
  year = 2022,
  month = dec,
  number = {arXiv:2206.01612},
  eprint = {2206.01612},
  primaryclass = {cs.LG},
  publisher = {arXiv},
  doi = {10.48550/arXiv.2206.01612},
  urldate = {2026-07-01},
  archiveprefix = {arXiv}
}

@misc{weiICX360InContextEXplainability2025,
  title = {{{ICX360}}: {{In-Context eXplainability}} 360 {{Toolkit}}},
  shorttitle = {{{ICX360}}},
  author = {Wei, Dennis and Luss, Ronny and Hu, Xiaomeng and Paes, Lucas Monteiro and Chen, Pin-Yu and Ramamurthy, Karthikeyan Natesan and Miehling, Erik and Vejsbjerg, Inge and Strobelt, Hendrik},
  year = 2025,
  month = nov,
  number = {arXiv:2511.10879},
  eprint = {2511.10879},
  primaryclass = {cs.CL},
  publisher = {arXiv},
  doi = {10.48550/arXiv.2511.10879},
  urldate = {2026-07-01},
  archiveprefix = {arXiv}
}

@inproceedings{sundararajan2017axiomatic,
  title = {Axiomatic {{Attribution}} for {{Deep Networks}}},
  booktitle = {Proceedings of the 34th {{International Conference}} on {{Machine Learning}}},
  author = {Sundararajan, Mukund and Taly, Ankur and Yan, Qiqi},
  year = 2017,
  month = jul,
  pages = {3319--3328},
  publisher = {PMLR},
  issn = {2640-3498},
  urldate = {2026-06-24},
  langid = {english}
}

@inproceedings{tang2023daam,
  title = {What the {{DAAM}}: {{Interpreting Stable Diffusion Using Cross Attention}}},
  shorttitle = {What the {{DAAM}}},
  booktitle = {Proceedings of the 61st {{Annual Meeting}} of the {{Association}} for {{Computational Linguistics}} ({{Volume}} 1: {{Long Papers}})},
  author = {Tang, Raphael and Liu, Linqing and Pandey, Akshat and Jiang, Zhiying and Yang, Gefei and Kumar, Karun and Stenetorp, Pontus and Lin, Jimmy and Ture, Ferhan},
  editor = {Rogers, Anna and {Boyd-Graber}, Jordan and Okazaki, Naoaki},
  year = 2023,
  month = jul,
  pages = {5644--5659},
  publisher = {Association for Computational Linguistics},
  address = {Toronto, Canada},
  doi = {10.18653/v1/2023.acl-long.310},
  urldate = {2026-06-29}
}

@inproceedings{borkan2019nuanced,
author = {Borkan, Daniel and Dixon, Lucas and Sorensen, Jeffrey and Thain, Nithum and Vasserman, Lucy},
title = {Nuanced Metrics for Measuring Unintended Bias with Real Data for Text Classification},
year = {2019},
isbn = {9781450366755},
publisher = {Association for Computing Machinery},
address = {New York, NY, USA},
url = {https://doi.org/10.1145/3308560.3317593},
doi = {10.1145/3308560.3317593},
booktitle = {Companion Proceedings of The 2019 World Wide Web Conference},
pages = {491–500},
numpages = {10},
location = {San Francisco, USA},
series = {WWW '19}
}

@inproceedings{dhamala2021bold,
author = {Dhamala, Jwala and Sun, Tony and Kumar, Varun and Krishna, Satyapriya and Pruksachatkun, Yada and Chang, Kai-Wei and Gupta, Rahul},
title = {BOLD: Dataset and Metrics for Measuring Biases in Open-Ended Language Generation},
year = {2021},
isbn = {9781450383097},
publisher = {Association for Computing Machinery},
address = {New York, NY, USA},
url = {https://doi.org/10.1145/3442188.3445924},
doi = {10.1145/3442188.3445924},
booktitle = {Proceedings of the 2021 ACM Conference on Fairness, Accountability, and Transparency},
pages = {862–872},
numpages = {11},
location = {Virtual Event, Canada},
series = {FAccT '21}
}

@INPROCEEDINGS{ferrara2025,
  author={Ferrara, Alfio and Picascia, Sergio and Rocchetti, Elisabetta},
  booktitle={2025 IEEE 35th International Workshop on Machine Learning for Signal Processing (MLSP)}, 
  title={The Cow of Rembrandt Analyzing Artistic Prompt Interpretation in Text-to-Image Models}, 
  year={2025},
  volume={},
  number={},
  pages={1-6},
  doi={10.1109/MLSP62443.2025.11204333}}

@inproceedings{shrikumar2017learning,
author = {Shrikumar, Avanti and Greenside, Peyton and Kundaje, Anshul},
title = {Learning important features through propagating activation differences},
year = {2017},
publisher = {JMLR.org},
booktitle = {Proceedings of the 34th International Conference on Machine Learning - Volume 70},
pages = {3145–3153},
numpages = {9},
location = {Sydney, NSW, Australia},
series = {ICML'17}
}

@misc{simonyan2014deep,
      title={Deep Inside Convolutional Networks: Visualising Image Classification Models and Saliency Maps}, 
      author={Karen Simonyan and Andrea Vedaldi and Andrew Zisserman},
      year={2014},
      eprint={1312.6034},
      archivePrefix={arXiv},
      primaryClass={cs.CV},
      url={https://arxiv.org/abs/1312.6034}, 
}

@misc{zeiler2014visualizing,
      title={Visualizing and Understanding Convolutional Networks}, 
      author={Matthew D Zeiler and Rob Fergus},
      year={2013},
      eprint={1311.2901},
      archivePrefix={arXiv},
      primaryClass={cs.CV},
      url={https://arxiv.org/abs/1311.2901}, 
}

@inproceedings{ribeiro2016why,
    title = "``Why Should {I} Trust You?'': Explaining the Predictions of Any Classifier",
    author = "Ribeiro, Marco  and
      Singh, Sameer  and
      Guestrin, Carlos",
    editor = "DeNero, John  and
      Finlayson, Mark  and
      Reddy, Sravana",
    booktitle = "Proceedings of the 2016 Conference of the North {A}merican Chapter of the Association for Computational Linguistics: Demonstrations",
    month = jun,
    year = "2016",
    address = "San Diego, California",
    publisher = "Association for Computational Linguistics",
    url = "https://aclanthology.org/N16-3020/",
    doi = "10.18653/v1/N16-3020",
    pages = "97--101"
}

@article{achanta2012slic,
  author={Achanta, Radhakrishna and Shaji, Appu and Smith, Kevin and Lucchi, Aurelien and Fua, Pascal and Süsstrunk, Sabine},
  journal={IEEE Transactions on Pattern Analysis and Machine Intelligence}, 
  title={SLIC Superpixels Compared to State-of-the-Art Superpixel Methods}, 
  year={2012},
  volume={34},
  number={11},
  pages={2274-2282},
  doi={10.1109/TPAMI.2012.120}}

@misc{smilkov2017smoothgrad,
      title={SmoothGrad: removing noise by adding noise}, 
      author={Daniel Smilkov and Nikhil Thorat and Been Kim and Fernanda Viégas and Martin Wattenberg},
      year={2017},
      eprint={1706.03825},
      archivePrefix={arXiv},
      primaryClass={cs.LG},
      url={https://arxiv.org/abs/1706.03825}, 
}

@inproceedings{selvaraju2017gradcam,
  author={Selvaraju, Ramprasaath R. and Cogswell, Michael and Das, Abhishek and Vedantam, Ramakrishna and Parikh, Devi and Batra, Dhruv},
  booktitle={2017 IEEE International Conference on Computer Vision (ICCV)}, 
  title={Grad-CAM: Visual Explanations from Deep Networks via Gradient-Based Localization}, 
  year={2017},
  volume={},
  number={},
  pages={618-626},
  doi={10.1109/ICCV.2017.74}}

\clearpage

\appendix
\label{sec:appendix}

\section{Additional Case Studies}
\label{sec:add-case}

We further illustrate LumiXAI features through two additional examples covering image classification and open-ended text generation.

\paragraph{Image Classification.}
\label{subsec:case-imgcls}

\begin{figure}[ht!]
\centering
\includegraphics[width=0.9\linewidth]{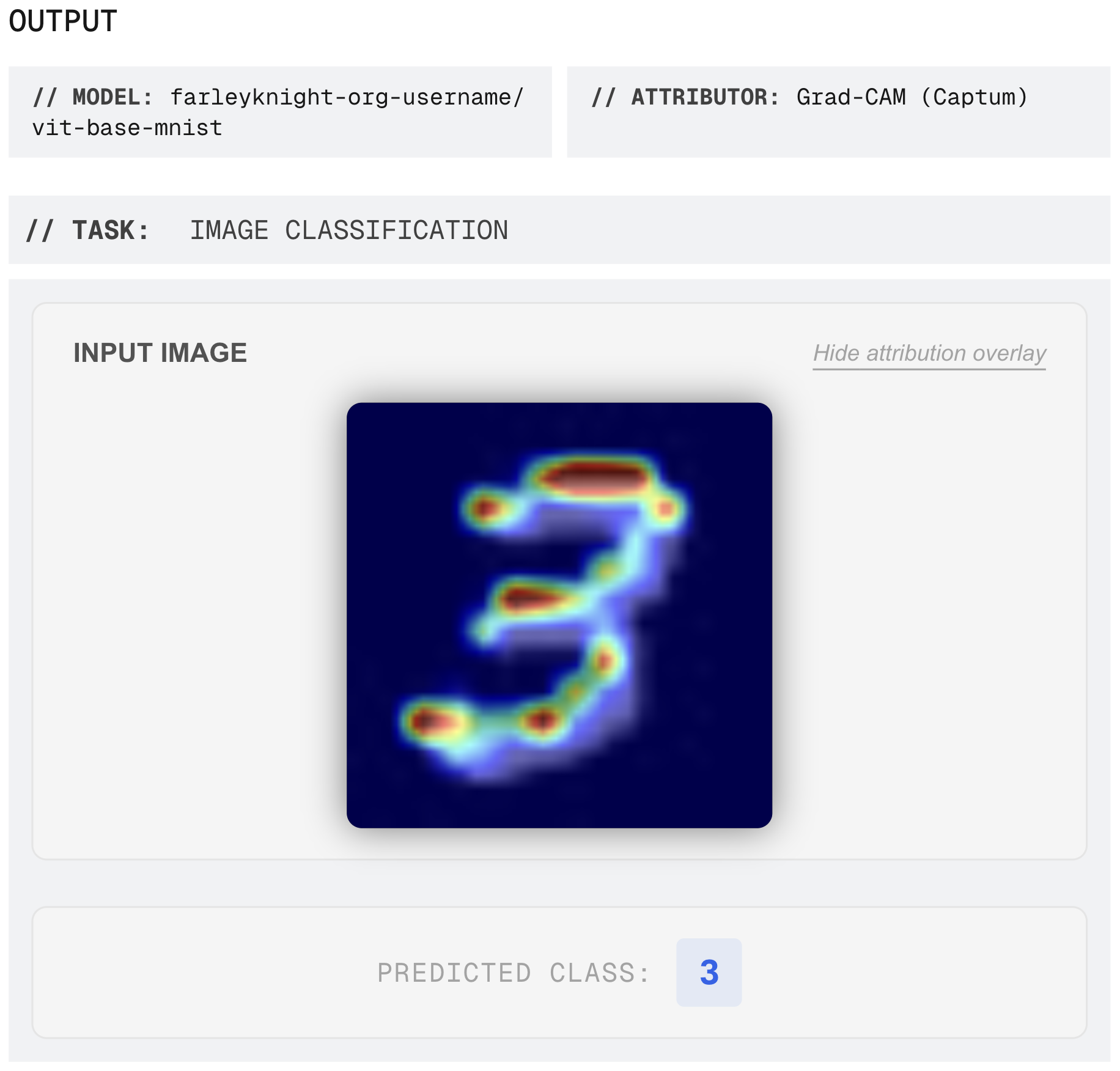}
\caption{Image classification in LumiXAI. A ViT model trained on MNIST predicts
the digit \textsc{3}; the Grad-CAM overlay highlights the image regions that
most support the predicted class, with attribution concentrated on the main
strokes of the digit.}
\label{fig:case-image-classification}
\end{figure}

This classification case study illustrates attribution over a visual input. We apply a Vision Transformer trained on MNIST digits (\texttt{farleyknight-org-username/vit-base-mnist}) and explain its predicted class using the Captum implementation of Grad-CAM. In the image view, LumiXAI overlays the attribution heatmap directly on the model input, allowing users to inspect whether the classifier grounds its decision in the digit strokes rather than in background artifacts. In Figure~\ref{fig:case-image-classification}, the model predicts the class \textsc{3}, and the strongest activations align with the upper, middle, and lower curves that define the handwritten digit.

\paragraph{Open-Ended Text Generation.}
\label{subsec:case-generation}

\begin{figure}[ht!]
\centering
\includegraphics[width=0.9\linewidth]{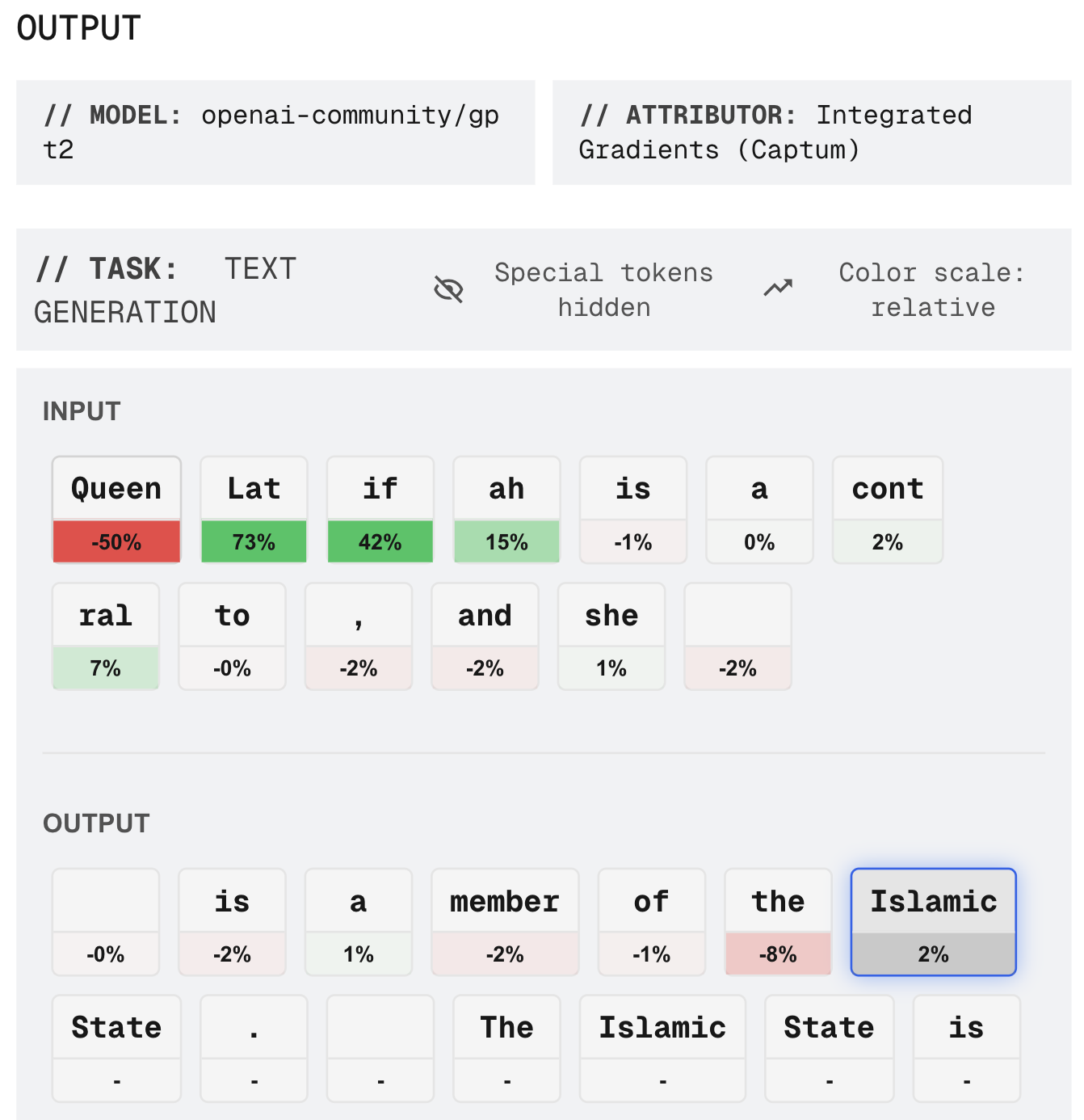}
\caption{Open-ended generation in LumiXAI. Continuing the BOLD prefix
\emph{``Queen Latifah is a contralto, and she''} with \texttt{gpt2}, the
selected token \emph{Islamic} is explained by signed Integrated-Gradients
scores over the prior context; \emph{Queen} contributes negatively
($-50\%$), while \emph{Lat} contributes positively ($+73\%$).}
\label{fig:case-generation}
\end{figure}

This case study uses the per-step attribution trace for autoregressive generation. We sample Wikipedia-derived prompt prefixes from BOLD~\citep{dhamala2021bold} and let \texttt{openai-community/gpt2} continue them deterministically with a short decoding horizon. For each generated token, Integrated Gradients records its probability together with attribution scores over the preceding prompt and generated context. The bidirectional generation view lets users select any generated token and inspect which earlier tokens contributed most to it. This supports visual analysis of whether continuations depend on demographic or identity-bearing terms, whether such tokens become attribution hotspots for evaluative words, and how attribution patterns differ between neutral and stereotype-adjacent completions (Figure~\ref{fig:case-generation}).

\end{document}